\documentclass[twocolumn]{aastex63}
\usepackage{rotating}
\usepackage{graphicx}
\usepackage{amssymb}
\usepackage{amsmath}
\usepackage{hyperref}
\usepackage{lineno}


\def\gtrsim{\mathrel{\hbox{\rlap{\hbox{\lower4pt\hbox{$\sim$}}}\hbox{$>$}}}}
\def\lesssim{\mathrel{\hbox{\rlap{\hbox{\lower4pt\hbox{$\sim$}}}\hbox{$<$}}}}
\def\farcs{\hbox{$.\!\!^{\prime\prime}$}}

\begin{document}

\title{
Probing the spectrum of the magnetar 4U\,0142+61 with JWST}
\author[0000-0002-8548-482X]{Jeremy Hare}
\affiliation{NASA Goddard Space Flight Center, Greenbelt, Maryland, 20771, USA}
\affiliation{Center for Research and Exploration in Space Science and Technology, NASA/GSFC, Greenbelt, Maryland 20771, USA}
\affiliation{The Catholic University of America, 620 Michigan Ave., N.E. Washington, DC 20064, USA}
\author[0000-0002-7481-5259]{George G. Pavlov}
 \affiliation{Department of Astronomy \& Astrophysics, Pennsylvania State University, 525 Davey Lab, University Park, PA 16802, USA}
 \author[0000-0003-2317-9747]{Bettina Posselt}
 \affiliation{Department of Astrophysics, University of Oxford, Denys Wilkinson Building, Keble Road, Oxford OX1 3RH, UK}
 \affiliation{Department of Astronomy \& Astrophysics, Pennsylvania State University, 525 Davey Lab, University Park, PA 16802, USA}
 \author[0000-0002-6447-4251]{Oleg Kargaltsev}
 \affiliation{Department of Physics, The George Washington University, 725 21st St. NW, Washington, DC 20052}
\author[0000-0001-7380-3144]{Tea Temim}
\affiliation{Princeton University, 4 Ivy Lane, Princeton, NJ 08544, USA}
\author[0000-0001-6169-2731]{Steven Chen}
\affiliation{Department of Physics, The George Washington University, 725 21st St. NW, Washington, DC 20052}

 \email{jeremy.hare@nasa.gov}

\begin{abstract}
JWST observed the magnetar  4U\,0142+61 with the MIRI and NIRCam instruments within a 77 min time interval on 2022 September 20--21.
The low-resolution MIRI spectrum and NIRCam photometry show that the spectrum in the wavelength range 1.4--11 $\mu$m range can be satisfactorily described  by an absorbed power-law model, $f_{\nu}\propto \nu^{-\alpha}$, with a spectral slope 
$\alpha =0.96\pm0.02$,
interstellar extinction $A_V=
3.9\pm0.2$, and normalization $f_0 = 59.4\pm 0.5$ $\mu$Jy at $\lambda = 8$ $\mu$m.
These observations do not support the passive disk model proposed by  \cite{2006Natur.440..772W}, based on the 
Spitzer 
photometry, which was interpreted as evidence for a fallback disk from debris formed during the supernova explosion. We suggest a nonthermal origin for this emission and source variability as the most likely cause of discrepancies between the JWST data and other IR-optical observing campaigns. However,  we cannot firmly exclude the presence of a large  disk with a different dependence of the effective disk temperature on distance from the magnetar.  Comparison with the power-law fit to the hard X-ray spectrum above 10 keV, measured by NuSTAR contemporaneously with JWST, shows that the X-ray spectrum
is significantly harder. This may imply that 
the X-ray and IR nonthermal emission come from different sites in the magnetosphere of the magnetar.

\end{abstract}

\section{Introduction}
Magnetars are neutron stars with extremely large inferred magnetic fields ($B\sim 10^{14}$--$10^{15}$ G) and long spin periods ($P=1-12$ s; \citealt{2014ApJS..212....6O}\footnote{See the McGill magnetar catalog at \url{http://www.physics.mcgill.ca/~pulsar/magnetar/main.html}}). Their persistent X-ray luminosities, 
substantially exceed their spin-down energy loss rates, 
suggesting that 
magnetar emission is powered by their strong magnetic fields. Soft X-ray (0.5--10 keV) spectra of magnetars are often described by a two-component, blackbody (BB) + power-law (PL), model with $kT\sim0.4$–0.6 keV, and spectral slope $\alpha \sim 1$--3
(energy flux density $f_{\nu}\propto \nu^{-\alpha}$). Magnetars have been also detected in hard X-rays (10--100 keV), where their spectra show an upturn ($\nu f_\nu$ increases with increasing photon energy).
For several magnetars observed multiple times in the NIR, variability has been reported (e.g., \citealt{2005ApJ...627..376D,2004A&A...416.1037H}).

Although the isolated and non-accreting magnetar interpretation is widely accepted, other hypotheses have also been discussed. One alternative explanation attributes the quiescent emission of magnetars to low-rate accretion, possibly from a ``fallback'' disk (e.g., \citealt{2000ApJ...543..368C,2000ApJ...541..344P,2003ApJ...599..450E}). Fallback in supernovae occurs when supernova ejecta is accreted by the newly formed compact object \citep{1971ApJ...163..221C,1988slmc.proc..341C,1989ApJ...346..847C}. This would naturally explain the long magnetar spin periods but can hardly explain
the powerful outbursts and high-energy tails
(but see, e.g., \citealt{2014A&A...562A..62K}, \citealt{2015MNRAS.454.3366Z}).
However, even if the fallback disks are not responsible for
all the X-ray behavior of magnetars, they could still exist and manifest themselves in different
ways and/or at other wavelengths.

Being one of the brightest 
 magnetars, the ``anomalous X-ray pulsar''
  4U\,0142+61 
  was detected in X-rays with Uhuru
  \citep{1978ApJS...38..357F}, and its X-ray period, $P\simeq 8.7$ s, was
  discovered with EXOSAT \citep{1994ApJ...433L..25I}. 
 The energy loss rate (or spin-down power) of the magnetar, $\dot{E}=1.2\times 10^{32}$ erg s$^{-1}$, is about 3 orders of magnitude lower than its quiescent X-ray luminosity, $L_{\rm 2-10\,keV} \approx 1\times 10^{35}$ erg s$^{-1}$ at the the most likely distance $d=3.6$ kpc.
  4U\,0142+61
  exhibits a typical magnetar-like X-ray spectrum
 and energy-dependent pulse profiles with a pulsed
  fraction of $\approx 4\%$ at energies of a few keV
\citep{2006csxs.book..547W}. On 2007 February 8   
4U\,0142+61 produced a 
strong outburst detected  
in  RXTE monitoring data \citep{2008AIPC..983..234G}.
During the outburst, the peak flux in the 2--60 keV band exceeded
the quiescent level by a factor of 500--1000.
Recent X-ray polarization measurements \citep{2022Sci...378..646T} are consistent with a model in which thermal radiation from the magnetar surface is reprocessed by scattering off charged particles in the magnetosphere.

At longer wavelengths, an  optical source having peculiar colors was detected in Keck observations by
\cite{2000Natur.408..689H} at the X-ray position of 4U\,0142+61, and it was the
first identified optical counterpart of a magnetar. Its faintness
($R\approx 25$ in 1999 September)
argued against an active accretion disk as the source of the observed
optical emission.
In addition, strong optical pulsations, with a
pulsed fraction
  of $\approx28$\%  
(\citealt{2002Natur.417..527K,2005MNRAS.363..609D}), were detected at the X-ray period, 
suggesting that
 most of the optical emission
originates in the magnetar's magnetosphere 
(but see \citealt{2004ApJ...605..840E,2007ApJ...657..441E}).  
The optical and X-ray pulse profiles show
similar morphologies and
are approximately in phase with each other. 
The NIR emission  of 4U\,0142+61 was first investigated by \cite{2004A&A...416.1037H} who found the $K$ ($K_s$) magnitudes to be in 19.6--20 mag range from 3 observations with NIRC on Keck I.  \cite{2004A&A...416.1037H} 
inferred significant variability in NIR,
later confirmed by \cite{2006ApJ...652..576D} who re-analyzed all the
NIR-optical data. However, the optical-NIR variability was later disputed by \cite{2016MNRAS.458L.114M} who re-analyzed some of the archival data and claimed that the source was not variable in optical data over a 12 year timespan. 4U 0142+61 was also observed with HST WFC3 in NIR  in 2018  and detected at 
$22.64\pm0.01$ and $22.07\pm 0.01$ mag (in the AB magnitude system) in  F125W and F160W filters, respectively \citep{2022MNRAS.512.6093C}, which differs by $<2\sigma$ from the fluxes measured from previous ground-based observations in $J$ and $H$ filters.

 As part of a systematic search for supernova debris disks
 around young NSs,
predicted by current supernova models \citep{2003ApJ...591..288H},
 4U\,0142+61's mid-IR counterpart was detected with  Spitzer IRAC
in the 4.5\,$\mu$m and 8.0\,$\mu$m bands by \cite{2006Natur.440..772W}. From these  
measurements and earlier optical-NIR observations,
which were not 
contemporaneous with the Spitzer observations, \cite{2006Natur.440..772W} inferred an IR excess
and concluded that this emission emerges from a  
 passive, non-accreting disk, illuminated by the magnetar's X-rays. After correcting for the interstellar reddening
 (assuming $A_V=3.5$), \cite{2006Natur.440..772W}  found
 that the intrinsic IR spectrum was consistent with
a 920 K blackbody or, even better, with a multi-temperature
 (700$-$1200\,K) thermal disk model.
They proposed that the disk is a dusty remnant of fallback
 material from the supernova that created the magnetar, which would make it the first supernova debris disk directly detected around
 a young NS.  Four additional IRAC observations, taken 2--3 weeks after after a burst on 2007 February 7,   showed virtually the same fluxes in the 4.5 um and 8.0 um bands \citep{2008ApJ...675..695W}.
 
 In a subsequent paper, \cite{2008AIPC..983..274W} analyzed {\sl Spitzer} IRS data and  argued that the spectrum may contain a 9.7\,$\mu$m emission feature, which could be a signature of dust emission from silicate grains. From a Spitzer MIPS observation, \citet{2008AIPC..983..274W} also reported an upper limit at 24\,$\mu$m, showing a turndown of the magnetar's spectrum at longer wavelengths.
 
 These 
 findings motivated us
 to propose contemporaneous  low-resolution spectroscopy in IR and photometry in NIR with JWST,
 supplemented by Swift-XRT and NuSTAR observations in X-rays \citep{2021jwst.prop.2635P}.  Section \ref{IR_obs_analysis} provides a description of  the MIRI, NIRCam, Swift-XRT, and NuSTAR observations and their analyses. We discuss our findings in Section \ref{discuss} and conclude with a summary in Section \ref{summary}.

\section{Observations and data analysis}
\label{IR_obs_analysis}
\subsection{JWST NIRCam imaging and photometry }

The Near Infrared Camera (NIRCam; \citealt{2023PASP..135b8001R}) was used to image the magnetar and measure 
its flux in two spectral bands.
The NIRCam observations were carried out 
on 2022 September 21 from 00:34:19 to 00:43:48 UTC.
The F250M 
and F140M filters
were employed in the long and short-wavelength NIRCam channels, using the NRCBLONG and NRCB1 detectors, respectively, in the SUB400P subarray configuration. The pivot wavelengths and bandwidths for these filters are listed in Appendix \ref{sec:append_A} and Figure~\ref{fig:jwst_spect} shows the wavelength dependencies of the filter throughputs.

We used the RAPID readout pattern, having seven total dithered integrations and 10 groups per integration. 
The effective scientific exposure time in each of the F250M and F140M filters is 115.9\,s. 
We use the pipeline-processed data (calibration software version 1.11.4; \citealt{2023zndo...8247246B}). 
In the drizzle process, a pixel shrinking of 1.0, pixel scale ratio of 1.0, and  inverse variance map (IVM) weighting scheme were used for the final resampling of the data. The nominal pixel sizes are 62.7\,mas and 30.7\,mas for the F250M and F140M images, respectively. 

Both the 250M and F140M images show a strongly nonuniform background,  dominated by the so-called 1/f noise patterns \citep{2020AJ....160..231S}. The details of the source photometry for these noisy images are described in Appendix \ref{sec:append_A}. The net source flux densities in the F250M and F140M filters are $f_\nu=14.5 \pm 0.4$\,$\mu$Jy and $5.1\pm 0.3$\,$\mu$Jy, respectively. The statistical uncertainties here and throughout the paper are reported at the $1\sigma$ level unless noted otherwise.

For the F250M filter, which has a larger field of view, we compared the radial profile of 4U 0142+61 with those of other (likely stellar) sources in the field and found them to be consistent with each other. Hence, we conclude that there is no indication of any extended NIR emission around the magnetar. The field of view for the F140M filter was smaller and no other sources were detected, leaving nothing to directly compare the radial profile to. The empirical PSF FWHM for the F250M (F140M) filter is 85 (48) mas, which constrains the angular size of the source to be $\lesssim0.1''$. At a distance of 3.6 kpc, the F250M limit corresponds to a physical limit on the size of a disk or IR wind-nebula of $\lesssim10^{16}$ cm or $\lesssim2\times10^{5}$ $R_{\odot}$.

\subsection{JWST MIRI spectroscopy with Low Resolution Spectrometer}
\subsubsection{Observation and data reduction}
\label{sec:lrs_data_reduction}
The Mid-Infrared Instrument (MIRI; \citealt{2015PASP..127..584R}) on JWST observed 4U\,0142+61 on 2022 September 20 starting at 23:26:56 UTC. We employed the MIRI Low Resolution Spectrometer (LRS; \citealt{2015PASP..127..623K}), which has a resolving power  ${\cal R}\equiv\lambda/\Delta\lambda \approx 24(\lambda/1\,\mu{\rm m}) -80$ in the 5--12 $\mu$m range.
A two-point dither ALONG SLIT NOD, which places the source at two different positions in the slit so that the sky background could be subtracted, was used.  The MIRI LRS observed 4U 0142+61 for 1975.8 s, and the data were readout using the FASTR1 mode. An acquisition image was obtained in the MIRI F560W filter (using the FASTGRPAVG readout) with an exposure time of 44.4~s.

We performed several different extractions of the 1D spectrum using the JWST pipeline to assess the impact of using non-default extraction parameters (e.g., extraction center and width, linear versus polynomial trace, different background regions). For instance, some MIRI LRS spectra have been found to be offset from the aperture position used by the standard pipeline, which can lead to poor extractions of the 1D spectrum (see, e.g., \citealt{2023ApJ...945L...2D}). These additional extractions did not exhibit any significant shift in the normalization or shape of the spectrum compared to the spectrum extracted by the standard pipeline. Therefore, we simply used the JWST pipeline extraction of the 1D spectrum for analysis, which was extracted using version 1.13.3 of the calibration pipeline 
\citep{2024_10463537}. We analyzed the  spectrum in the 5--11 $\mu$m wavelength range, 
since that is where the source dominates over the background. 

 Several issues remained with the spectrum after reducing it with the standard JWST pipeline. The first is that there was a hot pixel in one of the MIRI LRS observation nods that was not removed by the pipeline. The hot pixel appears as a very bright pixel in the 
 2D spectrum 
and as a single, bright but narrow line-like feature in the 1D
spectrum at 9.9\,$\mu$m. To remove this feature, we took the mean of the nearest 4 points (spanning $\sim$ 0.07$\mu$m), using the values of the two next longer and shorter wavelengths, and replaced the hot pixel value with this mean value. The second issue is that two points near 6.5 $\mu$m did not have any associated uncertainties,
so we adopted the error bars from neighboring points for these two points. 

Another issue is that 
the uncertainties of the spectral data points appear to be underestimated (e.g., neighboring values often fluctuate 
by a much larger amount than the pipeline-calculated 
uncertainties). Because of this error underestimation, one has to add hypothetical systematic errors to get acceptable spectral fits with reasonable models (see Section \ref{sec:jwst_spectral_fits}).

\subsubsection{
Modeling the JWST spectrum of 4U\,0142+61}
\label{sec:jwst_spectral_fits}
\begin{figure*}[t]
    \centering
    \includegraphics[scale=0.50]{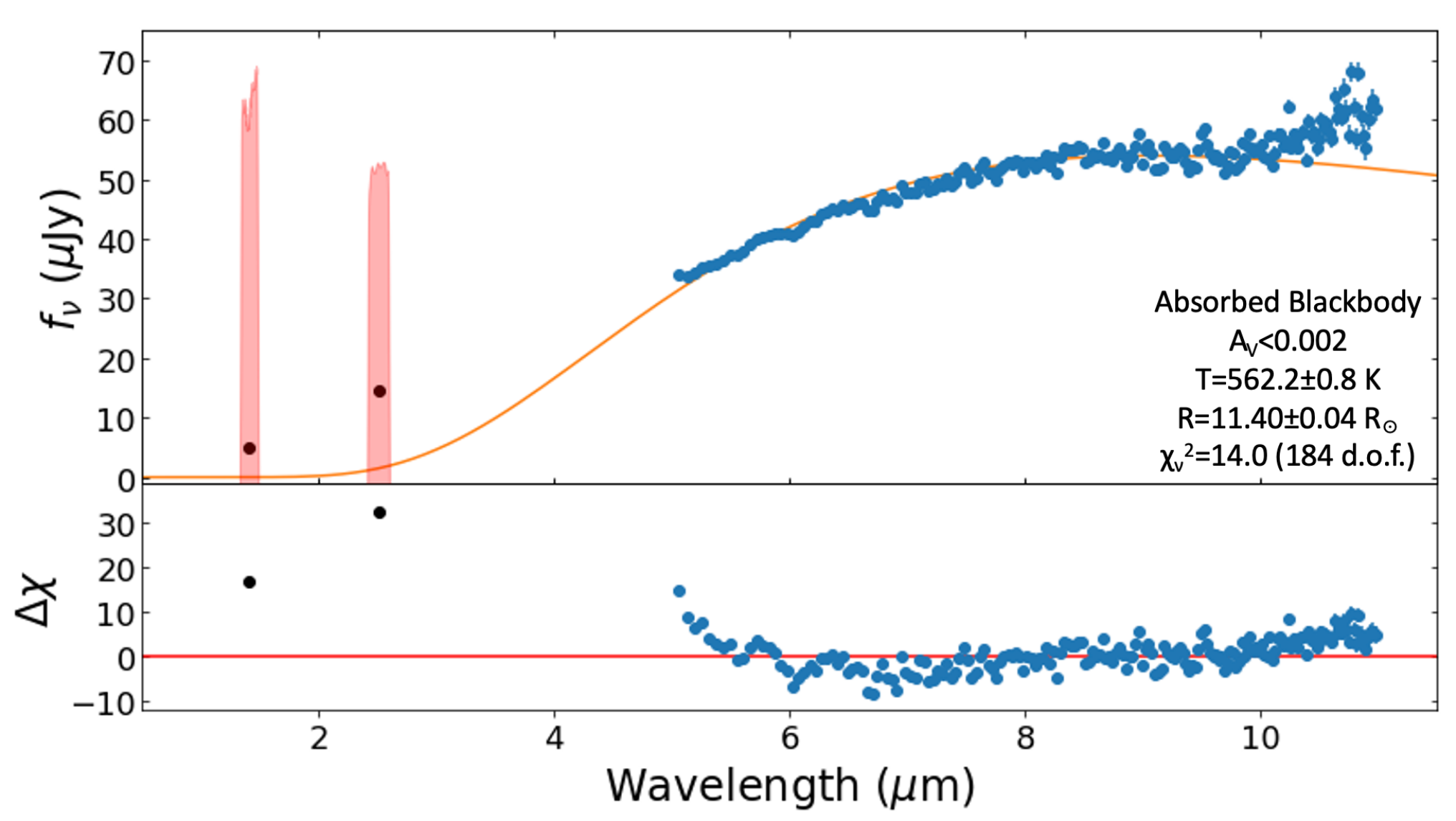}
    \includegraphics[scale=0.50]{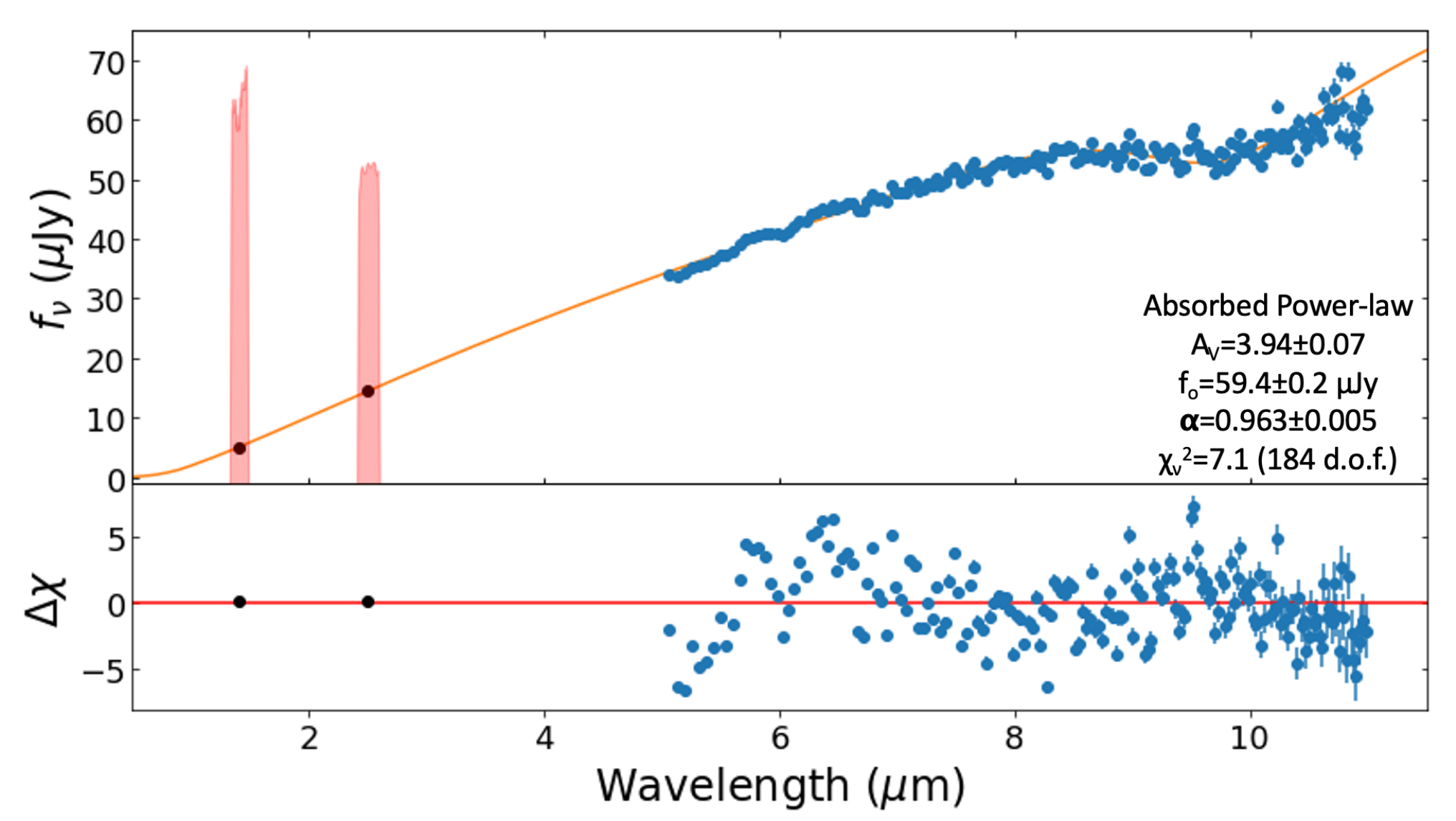}
    \caption{
    Spectral fits and fit residuals of the observed MIRI LRS spectrum (blue points) with absorbed blackbody (top) and power-law (bottom) models. 
The NIRCam F140M and F250M flux densities (black) are shown but not included in the fits. The semi-transparent red areas show wavelength dependencies of the throughputs of these filters. The residuals are defined as $\Delta \chi=$(data -- model)/error, where the values of `data` and `error' are supplied by the JWST calibration pipeline. 
}
\label{fig:jwst_spect}
\end{figure*}

\begin{figure}[t!]
    \centering
    \includegraphics[scale=0.4]{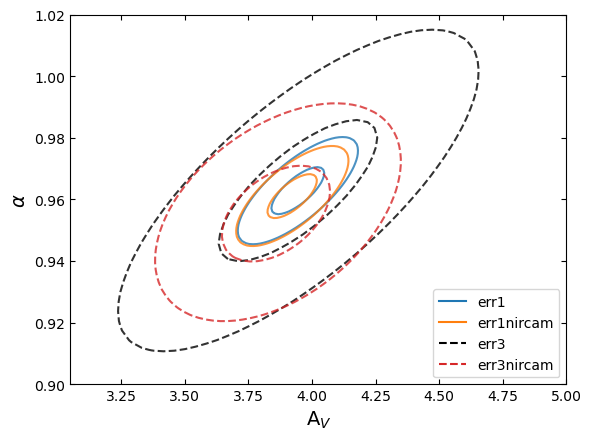}
    \caption{$1\sigma$ and $3\sigma$ confidence contours in the extinction - spectral slope plane for the absorbed 
    PL fits to the MIRI LRS only and MIRI LRS+NIRCam data. Two sets of contours are shown, one set shows the contours using the pipeline-produced MIRI LRS uncertainties (err1), while the other shows the  contours when inflating the MIRI LRS uncertainties by a factor of 3 (err3; see Section \ref{sec:jwst_spectral_fits}). Note that the NIRCam uncertainties are not inflated in the latter case. 
    }
    \label{fig:IR_cont}
\end{figure}

We start from fitting the MIRI LRS spectrum with two simple models.
Firstly, 
following \citet{2006Natur.440..772W}, we use an absorbed blackbody (BB) model 
to look for a thermal IR source,
such as a small disk with a nearly constant temperature.
To account for absorption, here and in other fits and estimates, we use
 the {\tt dust\_extinction} package \citep{2023ApJ...950...86G}, with the $V$ band extinction $A_V$ as a fitting parameter. 
We found that the best-fit BB model has a temperature $T=562.2\pm0.8$ K and an equivalent sphere radius $R=(11.40\pm0.04)R_{\odot}$ (at $d=3.6$ kpc), 
but the fit is clearly unacceptable (see Figure \ref{fig:jwst_spect}), with a large reduced chi-squared,  $\chi_\nu^2=14$ for $\nu=184$ degrees of freedom (dof).  
Additionally, the best-fit absorption in this model is $A_V=0$, which is inconsistent with the previously reported value of $A_V=3.5\pm0.4$ \citep{2006ApJ...652..576D}.
The best-fit BB model and residuals to the LRS spectrum are shown in the top panel of Figure \ref{fig:jwst_spect}. Note that the NIRCam points are not used in the fit (or 
$\chi^2$ calculation), but the extrapolation to the NIRCam wavelengths also show this model is incompatible with the data. Thus, we conclude that the 5--11 $\mu$m emission cannot be explained by a single-temperature thermal model.

The second model used is an absorbed power-law (PL) model, i.e., $f_\nu = f_0 (\lambda/\lambda_0)^\alpha 10^{-0.4 A_\lambda} = f_0 (\nu/\nu_0)^{-\alpha} 10^{-0.4 A_\nu}$. 
Fitting with this model yields an
extinction parameter $A_V=3.94\pm0.07$, spectral index $\alpha=0.963\pm0.005$, and normalization 
$f_0=59.4\pm0.2\,\mu$Jy, at $\lambda_0=8$ $\mu$m.
The fit is significantly better than  the absorbed BB model fit (see Figure \ref{fig:jwst_spect}), 
but $\chi_\nu^2=7.1$ for $\nu=184$  
is still too large to make it formally acceptable. 
This is likely due to the 
uncertainties  
being underestimated, particularly at shorter wavelengths. 
To make the fit formally acceptable, one has to significantly increase the pipeline-produced uncertainties.
For instance, we find that multiplying each of the pipeline-produced uncertainties by a factor of 3, 
gives a $\chi_\nu^2=0.79$.
This does not change the best-fit values but increases the uncertainties of the fitting parameters: $A_V=3.9\pm0.2$, 
$\alpha=0.963\pm0.015$, 
and 
$f_0=59.4\pm0.5\,\mu$Jy\footnote{We note that increasing the uncertainties for the BB model gives $\chi_\nu^2=1.55$, but there are still strong systematic residuals, particularly at short wavelengths, while the NIRCam points remain discrepant with this model.}.  

The best-fit PL model and residuals to the LRS spectrum (using the pipeline-produced uncertainties) are shown in the bottom panel of Figure \ref{fig:jwst_spect}. Again we note that the NIRCam points are not used in the fit (or 
$\chi^2$ calculation), but the extrapolation to the NIRCam wavelengths also shows good agreement with the data. The residuals of the PL fit show a hint of a spectral feature around 6 $\mu$m.  We, however, do not have a plausible interpretation of this feature and suspect it is due to calibration errors. 

In Figure \ref{fig:IR_cont} we  show the $1\sigma$ and $3\sigma$ 
confidence contours in the $A_V$-$\alpha$ plane, using both the pipeline-produced uncertainties (err1 in Figure \ref{fig:IR_cont}) and the uncertainties multiplied by a factor of 3 (err3 in Figure \ref{fig:IR_cont}). We also plot the confidence contours both including and excluding the NIRCam data points. The best-fit parameters do not change much with the inclusion of  the  NIRCam points.

We note that the pipeline-produced LRS spectrum, used in our fits, is oversampled -- 
the number of wavelength bins (187 in the 5--11 $\mu$m band) is larger than the number of resolution elements ($\sim 80$ in the same band), 
which means that the original bins can be combined into broader bins 
without a significant loss of resolution. 
We have checked, however, that a moderate binning (e.g., by a factor of 3) does not change the best-fit parameter values and only very slightly increases their uncertainties.

We also note that throughout the analysis of the MIRI LRS spectrum several calibration updates have been released and have occasionally caused shifts larger than the formal uncertainties in the best-fit parameters. For instance, using an earlier calibration version, we found best-fit parameters of $A_V=3.3\pm0.2$, spectral index $\alpha=1.034\pm0.013$, and normalization 
$f_0=55.6\pm0.4\,\mu$J with a similar 
$\chi_\nu^2=0.76$ (multiplying the data uncertainties by a factor of 3). This shift due to calibration is much larger than the 
uncertainties obtained for a given calibration version. However, the calibration of the LRS should be continually improving throughout the mission, so we use the best-fit values from the latest calibration for the remainder of this paper, but caution that a later calibration update may cause further changes in the best-fit parameters.

\subsection{X-ray observations}
To examine a contemporaneous spectral energy distribution (SED) from IR to hard X-ray, we carried out nearly simultaneous observations of the magnetar with the NuSTAR and Swift X-ray observatories.

\subsubsection{Neil Gehrels Swift X-ray Telescope (XRT)}
Swift-XRT observed 4U\,0142+61 starting at 01:11:36 UTC on 2022-09-21. This observation overlapped with the NuSTAR observation but occurred about 25 minutes after the end of the JWST observation. The source was observed using windowed timing (WT) mode and having an exposure time of 2.2 ks. 
The source spectrum was extracted using the online Build XRT products pipeline\footnote{\url{https://www.swift.ac.uk/user_objects/}} \citep{2009MNRAS.397.1177E}. We binned the spectrum to have a minimum signal-to-noise ratio of five and fit it in the 0.5--10 keV energy range.

\subsubsection{NuSTAR}
The Nuclear Spectroscopic Array (NuSTAR) 
observed 4U 0142+61 simultaneously with JWST for 40.8 ks on 2022-09-20 (ObsID 30801028002) with the observation starting at 23:01:09 UTC (i.e., about 30 minutes before the start of the JWST observation). The data were processed and reduced using HEASOFT version 6.31.1, with version 2.1.2 of the NuSTAR data analysis
pipeline using calibration version 20221115. The source spectrum was extracted from a $70''$ radius circle centered on the source. The background spectrum was extracted from a $70''$ circle placed in a source-free region on the same detector as the source. The source count rate was about 0.6 cts s$^{-1}$ for FPMA/B in the 3-78 keV energy range. The spectra were binned to have a minimum signal-to-noise ratio of five and fit in the 3-78 keV energy range.The X-ray spectra were fit with XSPEC version 12.13.0c \citep{1996ASPC..101...17A}. We used the {\tt tbabs} absorption model with {\tt wilms} abundances \citep{2000ApJ...542..914W}.

\subsection{X-ray Spectral Modelling}
\label{sec:xray_spec_model}
\begin{figure*}[t]
\centering
\includegraphics[scale=0.3]{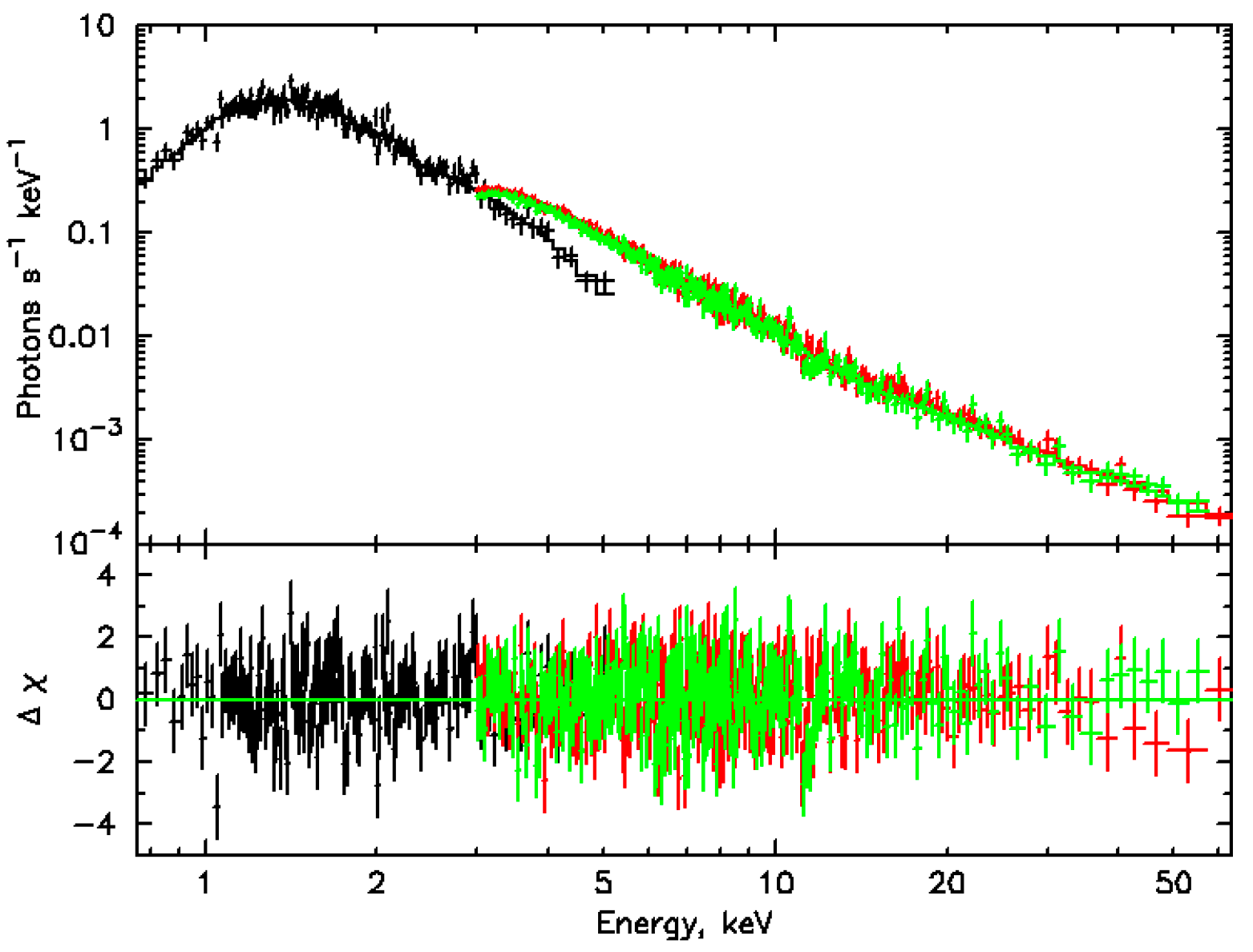}
\includegraphics[scale=0.3]{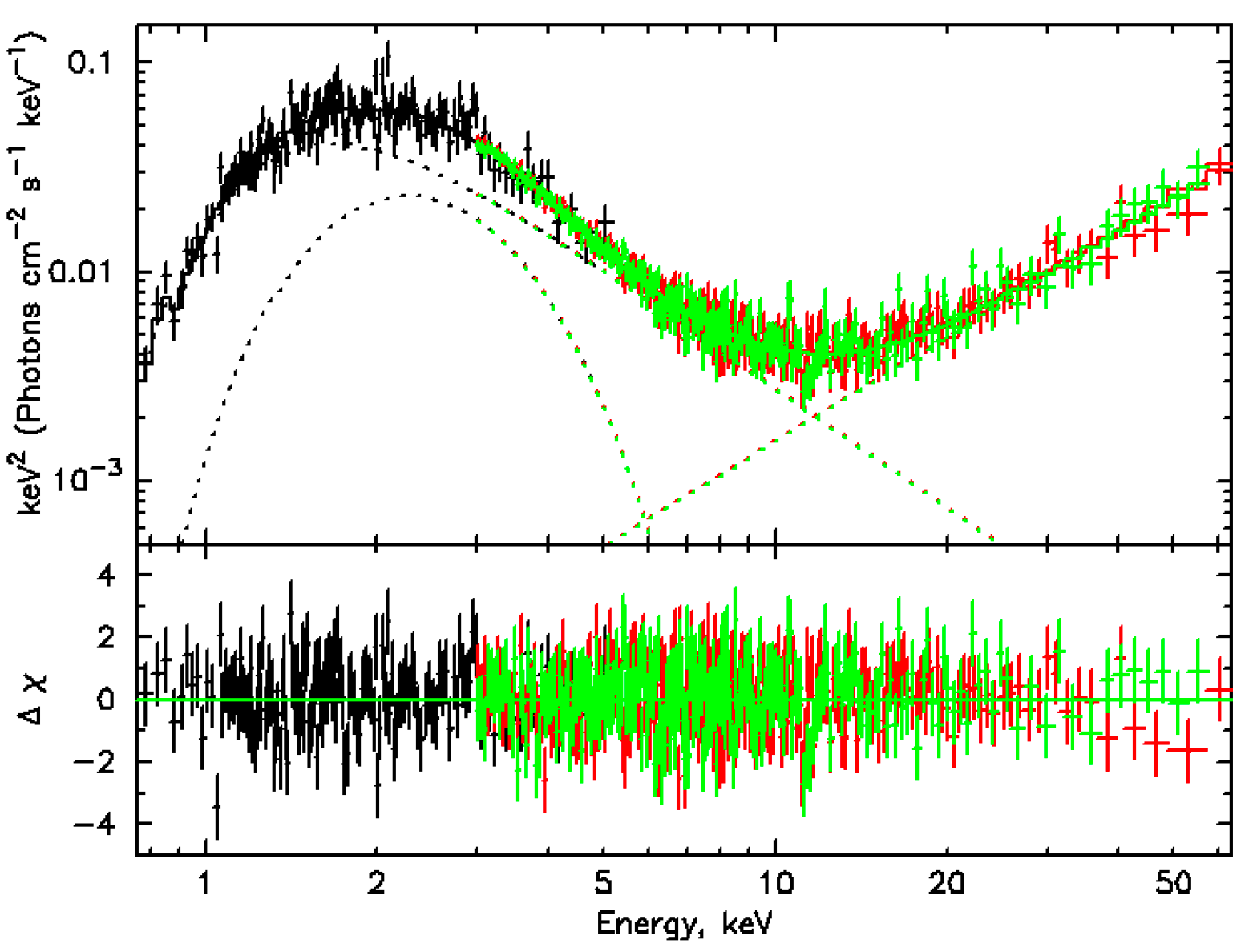}
\caption{{\it Left:} The observed Swift-XRT and NuSTAR X-ray spectra and best-fit 
BB+PL+PL model (model 1). The best-fit parameters are shown in Table \ref{tab:xray_models}. The residuals are shown in the bottom panel. {\it Right:} The same as left panel but for the unfolded $\nu f_{\nu}$ spectra. The model components are shown as dashed lines. 
}
\label{fig:xray_spect}
\end{figure*}
\begin{table*}
\begin{center}
\caption{Best-fit parameters for BB+PL+PL (model 1) and BB+BB+PL (model 2; see Section \ref{sec:xray_spec_model}) fit to the near simultaneous Swift+NuSTAR X-ray spectrum of 4U\,0142+61.
}
\begin{tabular}{ccccccccccc}
\hline\hline
   Model & Const. & $N_{\rm H}$ & $kT_1$ & Norm$_1$\tablenotemark{b}  & $\Gamma_s/kT_2$ & Norm$_2$\tablenotemark{b,c}  & $\Gamma_h$ & Norm$_3$\tablenotemark{c} & $F_{X}$\tablenotemark{d}  & $\chi^2$/dof \\
     \# & FPMA/FPMB\tablenotemark{a}   & $10^{22}$ cm$^{-2}$ & keV & &.../keV  & & & $10^{-5}$ & $10^{-10}$ c.g.s. &\\
    \hline
     1 & 1.00$^{+0.04}_{-0.03}$/0.99$^{+0.04}_{-0.03}$ & 1.58(8)  & 0.47(1)  & 138$^{+20}_{-18}$ & 3.90(8)  & 0.22(3) & 0.35(7) & 3.5$^{+0.8}_{-0.7}$ & 7.7(5) & 596.1/610 \\
     2 & 0.96(3)/0.95(3) & 0.54(4) & 0.464(8)
     & 302$^{+30}_{-26}$ & 1.09(3) & 1.2(2) & 0.83(6) & 20(3) & 1.68(8) & 634.0/610 \\
     \hline
\end{tabular}
\tablenotetext{a}{Normalization constant to account for differences in normalization due to calibration uncertainties between Swift-XRT and NuSTAR. The constant value is frozen to 1 for the Swift-XRT spectrum and fit for the NuSTAR spectra.}
\tablenotetext{b}{{\tt bbodyrad} normalizaion, $R^2_{\rm km}/d^2_{10}$, where $R_{\rm km}$ is the source radius in km and $d_{10}$ is the source distance in units of 10 kpc.}
\tablenotetext{c}{Power-law normalization in photons keV$^{-1}$ cm$^{-2}$ s$^{-1}$ at 1 keV.}
\tablenotetext{d}{Unabsorbed X-ray flux in the 0.5-80 keV energy band.}
\label{tab:xray_models}
\end{center}
\end{table*}

The broadband X-ray spectrum of 4U 0142+61 has been studied in detail in many previous works (see e.g., \citealt{2007MNRAS.381..293R,2010ApJ...716.1345G,2011PASJ...63..387E,2015ApJ...808...32T}). It has been shown in these works that the soft part of the X-ray spectrum, $E \lesssim$ 10 keV, requires two spectral components to get an acceptable fit, while a third component (namely a hard PL) is required if energies $\gtrsim$ 10 keV are included in the fit. We jointly fit the Swift and NuSTAR spectrum of 4U\,0142+61 to determine whether or not the hard 
PL tail connects smoothly to the IR spectrum, thus may be produced by the same particle population. To fit these spectra, we use an empirical absorbed 
BB+PL+PL model (model 1 hereafter), multiplied by a constant (frozen to 1 for Swift-XRT), to account for any cross-calibration differences between Swift and NuSTAR FPMA/B (i.e., {\tt const$\times$tbabs$\times$(bbodyrad+pow+pow)} in XSPEC). We also fit an empirical 
BB+BB+PL (model 2 hereafter) also multiplied by a constant (i.e., {\tt const$\times$tbabs$\times$(bbodyrad+bbodyrad+pow)}), as the previous work by \cite{2015ApJ...808...32T} showed that it can fit the spectra equally well, but provides a significantly different photon index for the hard power-law tail.

We report the best-fit values for both models in Table~\ref{tab:xray_models}. Statistically acceptable fits were achieved for both models, although the 
BB+PL+PL model fits the data better than the 
BB+BB+PL model. Overall, the best-fit parameters for model 1 from our dataset agree with the values found by \cite{2015ApJ...808...32T}. However, we find higher temperatures for both thermal components in model 2 and a harder photon index (i.e., $\Gamma_h$=0.83 for our fits  versus 1.03 from \citealt{2015ApJ...808...32T}). 
These differences may be due to changes in the NuSTAR calibration between the NuSTAR observations and analyses. In any case, we are primarily focused on the hard 
PL component of the spectrum, as it is the most likely to connect to the IR. We find that the photon index 
of the hard power-law tail is $\Gamma_h=0.35\pm0.07$ ($\alpha_h=-0.65\pm0.07$) or $\Gamma_h=0.83\pm0.06$ ($\alpha_h=-0.17\pm0.06$) for model 1 and model 2, respectively. Figure \ref{fig:xray_spect} shows the best-fit for model 1, the fit residuals, and the unfolded spectrum and individual model components. 

\section{Discussion}
\label{discuss}
The JWST observations have shown that the spectrum of 4U\,0142 in the 1.4--11 $\mu$m wavelength range can be described by an absorbed PL model, $f_\nu\propto \nu^{-\alpha}$, with a slope $\alpha\approx 1$ and extinction $A_V\approx 3.9$. This spectrum corresponds the observed and dereddened fluxes $F_{1.4-11\,\mu{\rm m}} =\simeq 4.2\times 10^{-14}$ and $F_{1.4-11\,\mu{\rm m}}^{\rm dered} \simeq 4.7\times 10^{-14}$ erg cm$^{-2}$ s$^{-1}$, and ``isotropic luminosity'' $L_{1.4-11\,\mu{\rm m}} \equiv4\pi d^2 F_{1.4-11\,\mu{\rm m}}^{\rm dered} \simeq 7.3\times 10^{31} d_{3.6}^2$ erg s$^{-1}$, where $d_{3.6} = d/3.6\,{\rm kpc}$. 
This IR luminosity is close to the magnetar's spin-down power, $L_{1.4-11\,\mu{\rm m}} = 0.61 \dot{E} d_{3.6}^2$, but it is much lower than the quiescent X-ray luminosity, $L_{1.4-11\,\mu{\rm m}}/L_{\rm 0.5-80\,keV} 
\sim 6 \times 10^{-5}$ or $2.8\times 10^{-4}$ for the BB+PL+PL and BB+BB+PL X-ray spectral models, respectively.

In this section we compare the JWST results with the results of previous IR-optical observations and discuss the multi-wavelength spectrum of the magnetar and the nature of its IR-optical emission.
\subsection{Comparison with the Spitzer results}
\label{sec:spit_compare}
\begin{figure*}[ht]
\centering
\includegraphics[scale=0.5]{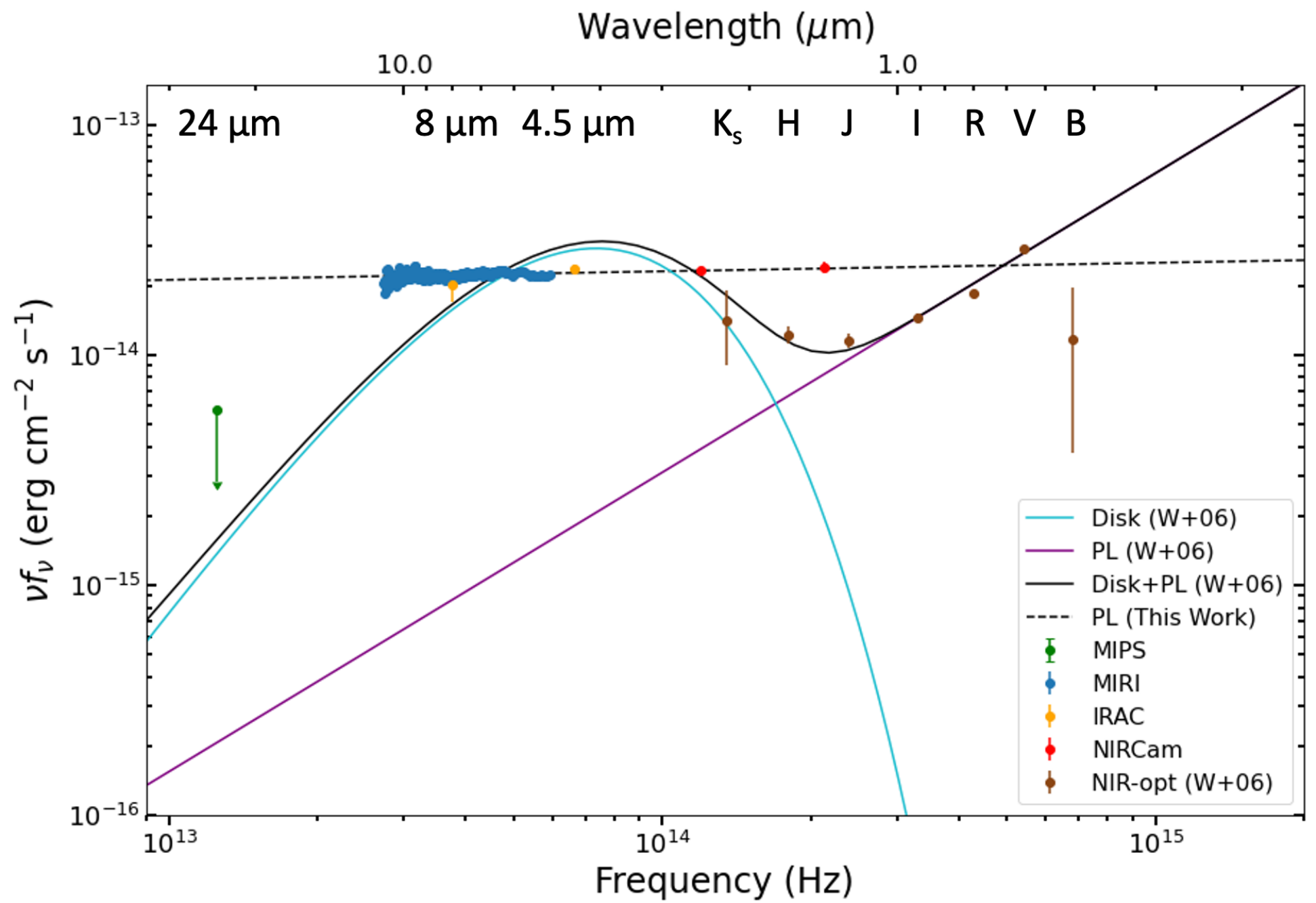}
\caption{Comparison of the dereddened (with $A_V=3.9$)  JWST  measurements and model spectra with the results from \cite{2006Natur.440..772W}.
The JWST MIRI LRS spectrum is plotted in blue, and the NIRCam points are plotted in red.
The 2005 Spitzer IRAC fluxes (\citealt{2008ApJ...675..695W}; see Section \ref{sec:spit_compare}) and MIPS upper limit \citep{2008AIPC..983..274W}  are shown in yellow and green, respectively. 
The brown points show NIR ($J,H,K_s$) and optical ($I,R,V,B$) photometry collected by \cite{2006Natur.440..772W} from previous publications. Note that the points from \cite{2006Natur.440..772W} and Spitzer were dereddened with $A_V=3.5$, but this difference causes only a small shift downward in the JWST points, and the large gap between the F140M and NIR points still remains.
The black dashed line is the best
PL fit 
($\alpha = 0.963$, $A_V=3.9$) 
to the MIRI LRS spectrum.
The solid black line shows the model adopted by \cite{2006Natur.440..772W}, which is the sum of the spectrum of a disk (cyan) with $T\propto r^{-3/7}$, $R_{\rm in} = 2.9 R_\odot$, $R_{\rm out} = 9.7 R_\odot$, 
and a PL with $\alpha = -0.3$ (violet).
The MIRI LRS and NIRCam points strongly disagree with this model.
}
\label{fig:wang_compare}
\end{figure*}
\cite{2006Natur.440..772W} reported on a Spitzer Infrared Array Camera (IRAC) observation of 4U\,0142+61, taken on
2005 January 17, however, \cite{2008ApJ...675..695W} re-analyzed these data and reported more accurate spectral fluxes (based on PSF fitting photometry instead of aperture photometry) $f_\nu=32.1\pm1.2$ and $48.8\pm 7.6$ $\mu$Jy at 4.5 and 8 $\mu$m, respectively. Spitzer also observed the source four times after it underwent an X-ray outburst on 2007 February 7. The IRAC observations spanned 2007 February 14-21 and measured average spectral fluxes, $f_\nu=32.1\pm 2.0$ and $59.8\pm 8.5$ $\mu$Jy at 4.5 and 8 $\mu$m, respectively \citep{2008ApJ...675..695W}, which are consistent with the observation from 2005. These points agree well with the JWST NIRCam photometry and MIRI spectrum, obtained almost 18 years later, with the largest offset between all of the Spitzer IRAC photometry and our PL fit being only about 1.5$\sigma$ for the 4.5$\mu$m flux measured from the 2005 observation (see Figures~\ref{fig:wang_compare}, \ref{fig:jwst_spit_compare}, and \ref{fig:jwst_spit_resid}).

Spitzer also observed the source with the Infrared Spectrograph (IRS) on 2006 January 22. The analysis of these data  was reported by \cite{2008AIPC..983..274W}, 
who discuss the
possible detection of a silicate emission feature at  9.7\,$\mu$m. Unfortunately, the Spitzer IRS spectrum is too noisy for a meaningful comparison with the MIRI LRS spectrum.  
However, we found no evidence of the 9.7\,$\mu$m emission feature in the LRS spectrum.

From the Spitzer observation with the Multiband Imaging Photometer (MIPS) 
on 2006 February 19, 
about a year after the Spitzer IRAC observations,
\cite{2008AIPC..983..274W} reported a $3\sigma$ upper-limit of 38 $\mu$Jy at 24 $\mu$m. \cite{2011AIPC.1379..152P} 
re-analyzed this MIPS observation
and found a  possible source at the position of 4U\,0142+61 in two independent, re-processed AORs (Astronomical Observation Requests), with a flux of $41$ $\mu$Jy, but it was only a $2\sigma$ significance detection. We use the latter measurement for this paper, since the analysis was performed using an updated Spitzer data reduction pipeline, and because it is close to the 3$\sigma$ upper-limit derived by \cite{2008AIPC..983..274W}\footnote{Note that the $3\sigma$ upper-limit from \cite{2011AIPC.1379..152P}  would be about 60$\mu$Jy}. This upper limit (or possible detection) from the MIPS observation strongly disagrees with the 
PL fit from the JWST data (see Figure~\ref{fig:wang_compare}). We find that a spectral break of about $\Delta\alpha \simeq 1.5$ (or $\Delta\alpha \simeq 1.0$, assuming a 60$\mu$Jy upper-limit) would be necessary to be consistent with the MIPS upper-limit assuming the break occurs at the long-wavelength end of the MIRI LRS spectrum (i.e., at 11$\mu$m).

\subsection{Previous optical-NIR observations}
\label{broadband_compare}
\begin{figure*}[t]
\centering
\includegraphics[scale=1.0]{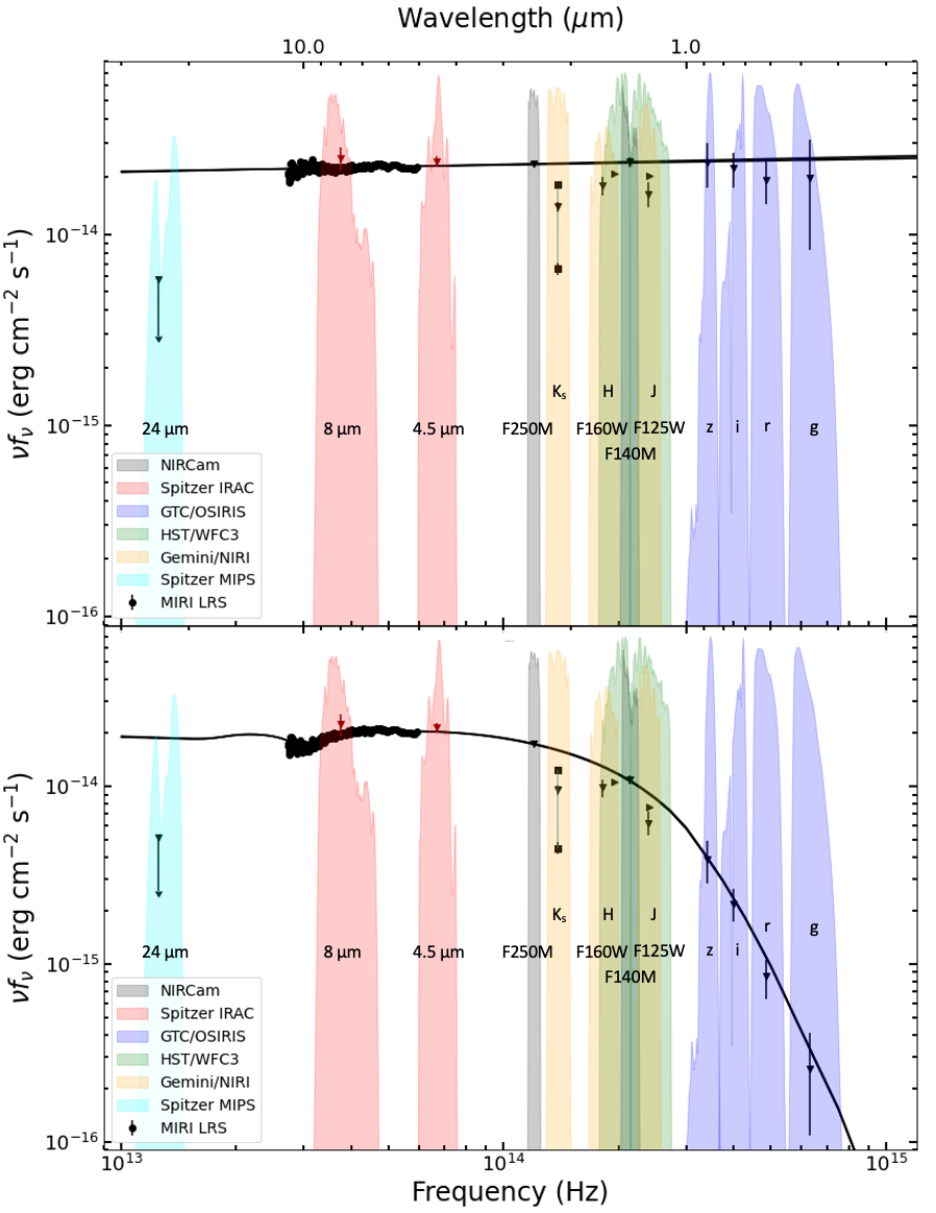}
\caption{Plot of the 
dereddened (top) and observed (bottom) broadband  spectral energy distribution of 4U\,0142+61. The points in the top panel were dereddened with the best-fit $A_V=3.9$ from the absorbed power-law model. The extrapolated best-fit absorbed PL model from the fit to the MIRI LRS spectrum is shown as a black line.  Note that the Spitzer IRAC fluxes shown here are the averages from the 2007 observations reported by \cite{2008ApJ...675..695W}, whereas those shown in Figure \ref{fig:wang_compare} were taken from the 2005 observation reported in  \cite{2008ApJ...675..695W}, thus are slightly different. The IR to optical photometry is taken from multiple publications (see Section  \ref{broadband_compare}). The black squares connected by a thin grey line in the $K_s$ band show the minimum and maximum $K_s$ band fluxes reported by \cite{2006ApJ...652..576D}. The right facing triangles show the HST points from \cite{2022MNRAS.512.6093C}.
}
\label{fig:jwst_spit_compare}
\end{figure*}
There have been numerous optical and NIR observations of 4U 0142+61 (e.g., \citealt{2000Natur.408..689H,2004A&A...416.1037H,2005MNRAS.363..609D,2005AdSpR..35.1177M,2006ApJ...652..576D,2016MNRAS.458L.114M,2022MNRAS.512.6093C}). \citet{2006Natur.440..772W} used published results from \cite{2004A&A...416.1037H} and \cite{2004IAUS..218..247I}, which included four optical ($B,V,R,I$) and three NIR ($J,H,K_s$) points (see Figure~\ref{fig:wang_compare}).
We also have gathered the results using different observations from those used in  \citet{2006Natur.440..772W} and plotted them 
in Figure~\ref{fig:jwst_spit_compare} (to avoid overcrowding the plots due to many data points measured in similar bands), together with the best PL fit of the JWST data.
In Figure~\ref{fig:jwst_spit_compare}, the 2007 Spitzer IRAC fluxes from \cite{2008ApJ...675..695W} are plotted. The $J, H$, and $K_s$ 
points, shown by filled triangles, were taken from the Gemini observations of 2004 November 2,
reported by \cite{2006ApJ...652..576D}. We added two square points that span the minimum and maximum $K$-band values reported by \cite{2006ApJ...652..576D}.
The 
F125W and F160W fluxes, in bands centered at 1.25 and 1.6 $\mu$m, 
were obtained by \cite{2022MNRAS.512.6093C} from HST WFC3/IR observations of 2018 January 1, and the optical
GTC/OSIRIS fluxes in the SDSS $z, r, i, g$ bands (which were all observed on the same night)
 were adopted from the 2013 observations reported by \cite{2016MNRAS.458L.114M}. We chose these datasets as they consist of observations across several filters which were obtained in the same observing run (i.e., nearly simultaneous for each individual dataset).

 \begin{figure*}[t]
\centering
\includegraphics[scale=0.5]{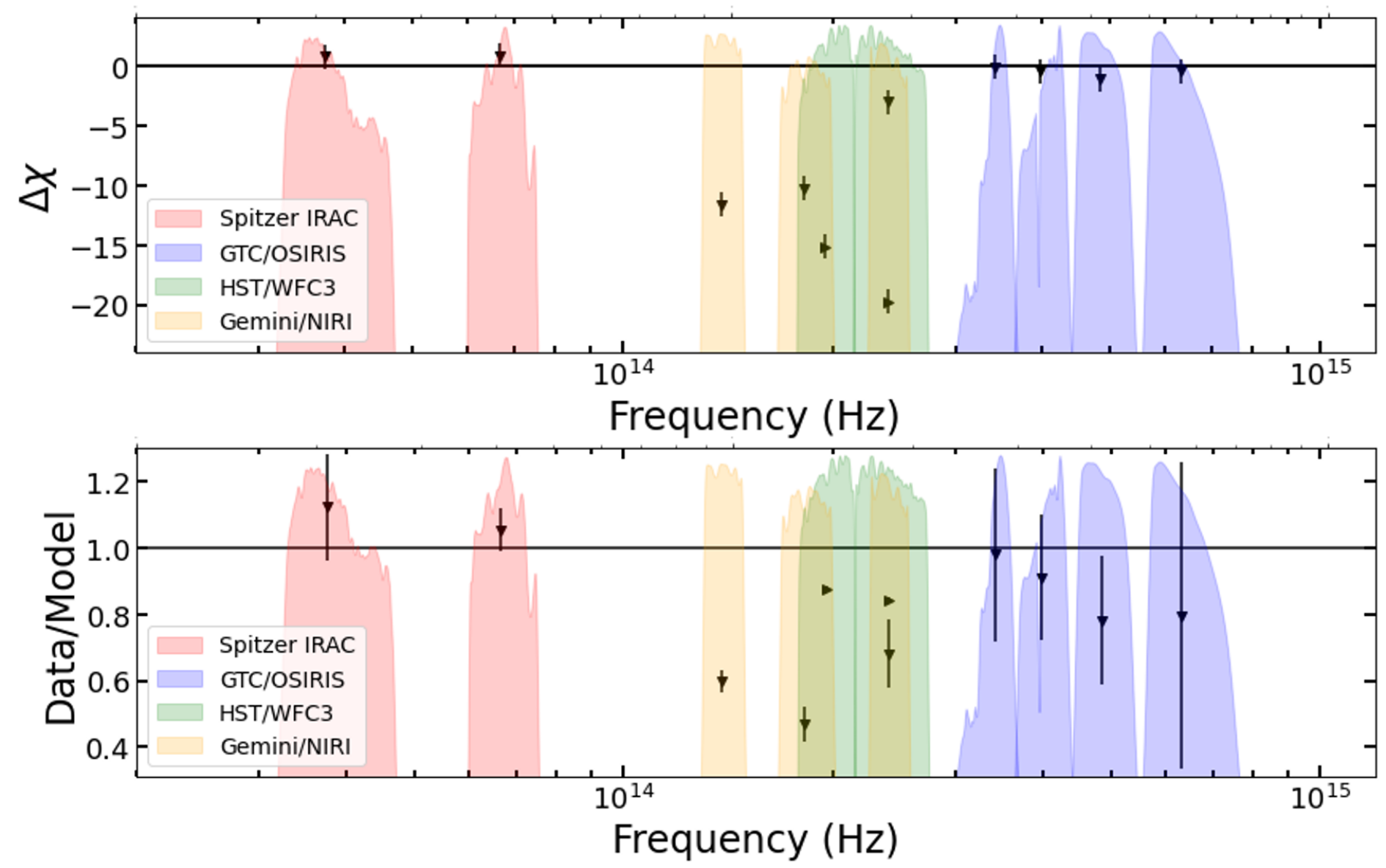}
\caption{ 
{\bf Top:} Discrepancies between the Spitzer IRAC, Gemini NIRI, HST WFC3, and GTC OSIRIS observations of 4U 0142+61 
and the best-fit absorbed 
PL fit to the JWST MIRI LRS spectrum, in units of $\sigma$. All points are shown as black triangles, with the exception of the HST points, which are shown as  right facing triangles. The error bars shown correspond to $\pm1\sigma$. Note that the Spitzer IRAC fluxes used here are the averages from the 2007 observations reported by \cite{2008ApJ...675..695W}.  {\bf Bottom}: The data-to-model ratio for the same data as shown in the top panel.
}
\label{fig:jwst_spit_resid}
\end{figure*}

As shown in Figures \ref{fig:wang_compare} and \ref{fig:jwst_spit_compare}, the results 
of these different observing campaigns exhibit 
offsets of different sizes 
from the
PL fit to the MIRI LRS data. 
Notably, the points in Figure \ref{fig:wang_compare} show a strong discrepancy with the high-frequency extension of the LRS PL fit, as well as with the results of
more recent NIR-optical observations. However, the strong discrepancies between the NIR data and LRS PL model still remain even in the more recent observations. The discrepancies, defined as $\Delta \chi=$(data-model)/$\sigma$ between the broadband data points from Figure \ref{fig:jwst_spit_compare} and the PL fit to the LRS spectrum, are shown in Figure \ref{fig:jwst_spit_resid}. 

The largest discrepancy between the model and photometry ($\approx15\sigma-25\sigma$) occurs in the HST bands, primarily due to the small uncertainties on the HST points. The second largest discrepancy occurs in the $K_s$ band, where the model overpredicts the $K_s$ flux by a factor of about 1.7 (or $\sim$11$\sigma$).  Additionally, the $H$ and $J$ band points lie below the model but are less discrepant than the $K_s$ band point (at about 10$\sigma$ and 5$\sigma$, respectively). The two HST points, which lie above the nearby $J$ and $H$ band points, are still about a factor of 1.2 smaller than flux predicted by the best-fit model to the LRS data. Surprisingly, the NIRCam points agree remarkably well with the best-fit PL model, even though these points were not included when fitting the model. 
One of these points lies a bit redward of the $K_s$ band point, while the other is between the $H$ and $J$ bands.
JWST is supposed to have an absolute flux calibration that is much better than $10\%$ (typically better than $2\%$), making calibration an unlikely cause for the discrepancies \citep{2022AJ....163..267G}. The optical points from \cite{2016MNRAS.458L.114M} also agree 
well with the extrapolated PL spectrum (as does their best-fit spectral index, $\alpha=0.8\pm0.5$, albeit with large uncertainties), but these points are the most reliant on the best-fit $A_V$, which has changed throughout the various calibration updates (see Section \ref{sec:jwst_spectral_fits}).  We discuss potential causes for these discrepancies, such as
variability and/or different emission sites for the IR and optical emission, in the following subsections.

\subsection{Variability}
\label{sec:variablility}
Since the IR-optical observations of the magnetar were taken at different epochs, an obvious explanation for the discrepancies between the MIRI LRS spectrum 
(and NIRCam photometry)
and previous broadband photometry is source variability (see Figure \ref{fig:jwst_spit_resid}).
Originally, \cite{2004A&A...416.1037H} reported variability at NIR wavelengths, particularly in the $K$ band, but noted that the source did not appear to be variable at optical wavelengths. $K$-band variability of greater than one magnitude, over a timescale of a few days,  was 
reported by \cite{2006ApJ...652..576D}. 
These authors 
also found that the source was possibly variable at optical wavelengths, by about half a magnitude in the $I$ band. 
On the other hand, \cite{2016MNRAS.458L.114M} reported a lack of variability in the optical $z$, $i$, $r$ and $g$ bands, but they did not comment on the possibility of NIR variability.  \cite{2022MNRAS.512.6093C} found that the NIR HST fluxes were consistent with the source being variable when compared to the NIR results of \cite{2004A&A...416.1037H} and \cite{2006ApJ...652..576D}. Additionally, using AKARI observations, \cite{2013IAUS..291..422K} found a 64\% lower flux at 4 $\mu$m compared to the Spitzer 4.5 $\mu$m flux and suggested that this discrepancy was due to variability.  

There is also the MIPS 24 $\mu$m upper limit, which is about a factor of three lower than the expected flux extrapolated from the best PL fit 
to the MIRI data. The factor of three discrepancy in flux 
is consistent with the extent of the variability reported by \cite{2006ApJ...652..576D} in the NIR (demonstrated in Figure~\ref{fig:jwst_spit_compare} for the $K_s$ band). Therefore, 
we cannot exclude the possibility that the 
mid-IR and NIR 
fluxes vary by about a factor of three, giving rise to the observed discrepancies. However, the good agreement of the Spitzer IRAC fluxes, from both \citealt{2006Natur.440..772W} and \citealt{2008ApJ...675..695W}, with the JWST LRS + NIRCam spectrum, observed 
17+ years later, as well as the lack of (or small scale) variability observed at optical wavelengths challenge this scenario. Additionally, Table 1 in \cite{2008ApJ...675..695W} shows that the IRAC flux of 4U 0142+61 did not vary over the week long observing campaign. 

It could also be the case that the IR and optical emission are produced by different 
mechanisms and/or at different sites near the magnetar, leading to variability in the 
IR that is not observed in the optical. 
However, there is 
a fair agreement between the optical observations and extrapolated best-fit 
PL model, suggesting that both emission components are produced by the same mechanism. Thus, the question of the origin and true extent of the possibly wavelength-dependent variability remains open.

\subsection{Possible contribution from a fallback disk}
\label{sec:MW_disk_fits}
Another 
explanation of the possibly different IR and optical spectra of 4U\,0142+61
is the presence of a fallback disk whose contribution to the observed emission may be different in different wavelength ranges. 
The discovery of optical pulsations \citep{2002Natur.417..527K,2005MNRAS.363..609D}, with a pulsed fraction of about 25\%--30\%, significantly higher than $\sim 4\%$ in X-rays, has led to perception that the optical emission comes from the magnetar's magnetosphere.
If true, it means that the hypothetical fallback disk 
would be more easily detected in the IR.
\cite{2006Natur.440..772W} noticed that the previously reported $V,R,I$ fluxes are consistent with a PL model with $\alpha = - 0.3$, at an assumed $A_V=3.5$. They used this PL 
in combination with a disk model 
to fit the IR + optical data
(see Figure \ref{fig:wang_compare}).
That model assumes that the optically thick disk is heated by the X-ray emission from the magnetar, the disk's local effective temperature decreases with 
radial distance from the magnetar as $T\propto r^{-3/7}$, and each point on the disk surface emits BB radiation with this temperature.  
\cite{2006Natur.440..772W}  found that the best-fit disk model had inner and outer disk temperatures $T_{\rm in} = 1200$ K and $T_{\rm out}=715$ K, and inner and outer disk radii $r_{\rm in}= 2.9 R_{\odot}$ and $r_{\rm out} = 9.7 R_{\odot}$, respectively, at $d=3.6$ kpc
and $\cos i = 0.5$, where $i$ is the disk inclination. Because of the small difference between the inner and outer temperatures, the disk spectrum resembles a BB spectrum with a temperature of 920 K. 

We reproduced the spectrum from \citet{2006Natur.440..772W} using a simple disk spectrum model given by 
Equation~(\ref{eq:multiT_BB_disk}). Figure  \ref{fig:wang_compare} shows the model disk+PL spectrum together with the results of
the optical/IR photometry (adopted from \citealt{2006Natur.440..772W} with updated IRAC fluxes from \citealt{2008ApJ...679.1443W}) and the JWST data.
It is immediately apparent that the JWST data 
are strongly inconsistent with this model. The MIRI spectrum does not have the anticipated curvature, and it overshoots the model at the long wavelength end. Additionally, the NIRCam F140M point greatly exceeds the anticipated model flux. Moreover, the most recent optical points from \cite{2016MNRAS.458L.114M} do not follow a PL with $\alpha=-0.3$, at least for the best-fit $A_V=3.9$ found from the PL fit of the MIRI LRS data, as we see from Figure~\ref{fig:jwst_spit_compare}. 

Although the PL+disk model used by \citet{2006Natur.440..772W} is inconsistent with the JWST data, it does not mean that another disk model, 
perhaps in combination with a PL component, cannot describe the 
IR or IR+optical data.
For instance, in the approximation of a multi-temperature BB flat disk, the flux density spectrum is
\begin{equation} 
f_\nu = \frac{2\pi \cos i}{d^2} \frac{2h\nu^3}{c^2} \int_{r_{\rm in}}^{r_{\rm out}} \frac{r\,dr}{\exp[h\nu/kT(r)] -1}\,.
\label{eq:multiT_BB_disk}
\end{equation}
If the radial dependence of the local effective temperature of 
a disk
obeys a PL, $T(r) = T_{\rm in} (r/r_{\rm in})^{-\beta} = T_{\rm out} (r/r_{\rm out})^{-\beta}$, then the 
disk spectrum is close to a PL with the slope $\alpha = 2/\beta -3$, i.e.,
\begin{equation}
    f_\nu \propto \nu^{3-2/\beta}\quad {\rm at}\quad (2/\beta - 1) kT_{\rm out} \ll h\nu \ll kT_{\rm in}\,.
    \label{eq:pl_disk}
\end{equation} 
Thus, the spectrum of a disk with a sufficiently large ratio of the inner and outer temperatures, $T_{\rm in}/T_{\rm out} \gg 2/\beta -1$ (i.e., the large ratio of the outer and inner radii, $r_{\rm out}/r_{\rm in} \gg (2/\beta -1)^{1/\beta}$) has a broad PL part. 
If we assume that the JWST PL spectrum is due to such a disk, then $\beta = 2/(\alpha +3) = 0.505$ for $\alpha=0.96$ (or $\beta = 0.5$ for $\alpha =1$). 

In a self-consistent disk model, the radial dependence of temperature should be derived from a balance between heating and emitted energies. In original models for a passive protostellar flat disk, illuminated by a central star, the dependence $T \propto r^{-3/4}$ was derived, 
which corresponds to $\alpha=-1/3$ (see \citealt{1987ApJ...312..788A}, and references therein). However, 
most protostellar disks exhibit flattish SEDs in the IR range, with $\alpha \approx 0.25$--1 (e.g., \citealt{1990AJ.....99..924B}), which corresponds to $\beta \approx 0.5$--0.6, 
similar to the slope we found for 4U\,0142+61. The difference between the 
observed and model temperature dependencies is likely due to some unrealistic assumptions in the original models. In particular, realistic disks should be flared rather than flat, and the local emission spectra are not BBs because the temperature of an irradiated passive disk should decrease inward (toward the disk mid-plane), and the disks become optically thin at large wavelengths \citep{1997ApJ...490..368C}. Therefore, the spectrum given by Equation (\ref{eq:multiT_BB_disk}) may be a crude (albeit useful) approximation. 

The entire spectrum of an optically thick, multi-temperature disk has three distinct regions.
For the flat disk spectrum given by Equation (\ref{eq:multiT_BB_disk}), these are a Rayleigh-Jeans tail at $h\nu \ll kT_{\rm out}$, where $f_\nu \propto \nu^2$,  a PL part given by Equation (\ref{eq:pl_disk}), and a high-frequency tail resembling a Wien spectrum, $f_\nu \propto \nu^2\exp(-h\nu/kT_{\rm in})$ at $h\nu \gg (2/\beta - 1) kT_{\rm in}$. One could assume that the high-frequency part might correspond to the cutoff of the optical spectrum between the $V$ and $B$ bands \citep{2004A&A...416.1037H}, seen in Figure \ref{fig:wang_compare}. However, such an interpretation requires a rather high temperature $T_{\rm in}  
\sim 7000$~K, which is 
well above the dust sublimation temperature for 
grains of any chemical composition (see e.g., \citealt{2002A&A...387..233K,2021MNRAS.501.3061T}). This means that it would be a gaseous disk. \citet{2007ApJ...657..441E} 
argue that the Spitzer NIR data, together with the data from preceding NIR-optical observations, can be interpreted as emission of a gaseous fallback disk with temperatures $T_{\rm in}\approx 6500$~K, $T_{\rm out} \approx 360$~K, and radii $R_{\rm in} \approx 7\times 10^9$ cm, $R_{\rm out} \approx 1.9\times 10^{12}$ cm. The $T(r)$ dependence provided in their Table 1 can be approximated by a PL, $T(r) \propto r^{-0.51}$, which leads to the spectral slope $\alpha = 0.92$, virtually coinciding with our measurement. \citet{2004ApJ...605..840E} and \citet{2007ApJ...657..441E} discuss how such a disk could be formed and how its emission could be strongly pulsed. 

Looking at Figure \ref{fig:jwst_spit_compare}, one might assume that the high-frequency cutoff of the disk spectrum corresponds to the $K$ band, which would not require such high values of $T_{\rm in}$. However, such an assumption does not look plausible because the $K$ band is in between the simultaneously observed NIRCam 2.5 $\mu$m and 1.4 $\mu$m points that lie on the extrapolation  of the LRS spectrum towards NIR. Although it is somewhat suspicious that all the $K$-band measurements are below the LRS spectrum, it is likely due
 to a chance coincidence.

One of 
appealing features of the 
disk interpretation of the (N)IR spectrum 
is that it 
could 
successfully capture the MIPS upper limit. Therefore, we also attempted to fit the disk model
given by Equation (\ref{eq:multiT_BB_disk}) to the MIRI LRS spectrum,
including the $2\sigma$ MIPS upper limit. 
We find, however, that this disk model is not able to account for 
 the MIPS upper limit 
while also matching the slope of the LRS spectrum, particularly the points at wavelengths longer than 8$\mu$m. Specifically, the rollover from the Rayleigh-Jeans tail to the flat part of the disk spectrum is not sharp enough to capture the longer wavelength portion of the 
LRS spectrum, so it undershoots the 
LRS data. Additionally, the slope of the flat part of this disk model is too soft and under-predicts the NIRCam points. 
If the MIPS point is ignored in the fit, then the best-fit disk model can successfully capture the MIRI LRS and NIRCam data, but it overpredicts the MIPS upper-limit by a factor of $\sim2$--3. 

It is important to note, however, that at long wavelengths the disk may become optically thin, so its flux at a given $\lambda$ becomes smaller by a factor of $1-\exp({-\tau_\lambda})$, where the disk's optical thickness $\tau_\lambda$ decreases with increasing wavelength (e.g., \citealt{1984ApJ...285...89D}). Therefore,
we cannot exclude the possibility that a 
fallback disk with 
$T_{\rm in}/T_{\rm out} \gtrsim 15$ (hence, $r_{\rm out}/r_{\rm in} \gtrsim 200$) significantly contributes to at least the IR spectrum of 4U\,0142+61.  

Overall, 
variability of nonthermal (magnetospheric) emission seems to be a more natural explanation of the MIPS upper limit discrepancy with the MIRI + NIRCam PL spectrum. However, the true cause of this discrepancy and the origin of the 4U\,0142+61's IR-optical emission can hardly be firmly established without 
nearly simultaneous observation with JWST and HST.

\subsection{Comparison between IR and X-ray spectra}
\begin{figure*}
\centering
\includegraphics[scale=0.52]{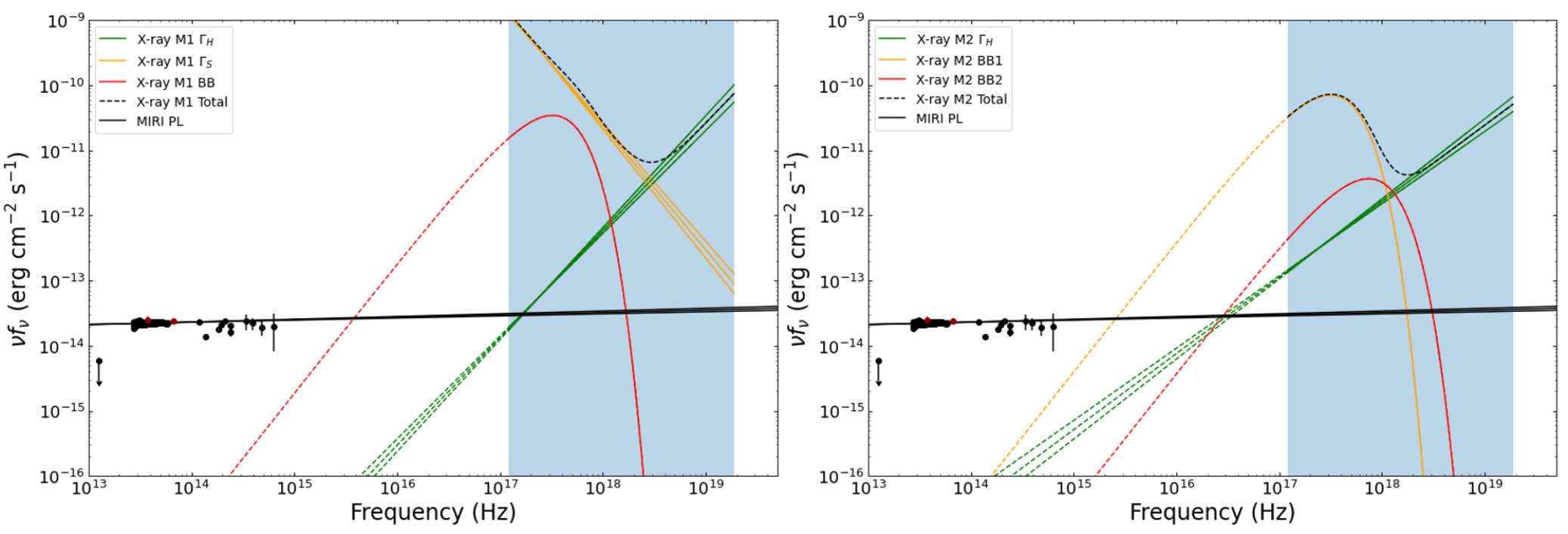} 
\caption{Broadband spectral energy distribution extending from IR to hard X-rays plotted with X-ray spectral model 1 (M1; left) and model 2 (M2; right, see Table \ref{tab:xray_models}). The broadband photometry is the same as plotted in Figure \ref{fig:jwst_spit_compare}. The black lines show the best-fit absorbed 
PL model (with uncertainties on the slope) to the MIRI LRS spectrum. The blue stripe shows the X-ray energy covered by Swift-XRT and NuSTAR. The model components on the left (right) are from the best-fit unabsorbed
BB+PL1+PL2 (BB1+BB2+PL) model (i.e., model 1 and 2, respectively) to the X-ray data and are shown in red, orange, and green respectively. The black dashed lines in both panels show the sum of the unabsorbed model components. This plot clearly shows that the hard  X-ray power-law component of the spectrum cannot continuously extend into the IR without a large spectral break for either model 1 or 2.}
\label{fig:broad_spect}
\end{figure*}
As we show in Section \ref{sec:jwst_spectral_fits}, the NIRCam and MIRI LRS data in the $\lambda=1.4$--11 $\mu$m band are well fit by an absorbed PL model with $\alpha\simeq0.96$. Since PL spectra are typical for emission of relativistic particles, this might suggest that the observed IR spectrum comes from
the magnetosphere of 4U\,0142+61 and can be connected with a nonthermal component of the X-ray spectrum, particularly, with the hard X-ray component, as suggested by \citet{2016MNRAS.458L.114M}.
However, the slopes of the IR and hard X-ray spectra are quite different, i.e., they cannot be connected without a large spectral break, $\Delta\alpha \simeq 1.6$ or $\Delta\alpha \simeq 1.1$ for models 1 and 2, respectively (see Figure \ref{fig:broad_spect}).
This suggests that 
even if the IR originates from the magnetosphere, the IR and hard X-ray emissions either come from two separate particle populations or are produced by different emission mechanisms. 
Future NIRCam timing observations can help to solidify the location of these IR/NIR emitting particles, the emission mechanism, and, depending on the IR spectrum at longer wavelengths, the strength of IR/NIR pulsations. Additionally, measuring the phase-shifts between the (N)IR and X-ray emission could help to determine the relative locations of the emission sites.

\section{Summary and conclusions}
\label{summary}
The JWST MIRI and NIRCam observations of the brightest magnetar 4U\,0142+61, supplemented by Swift-XRT and NuSTAR observations in X-rays, have shown the following.
\begin{itemize}
    \item The spectrum extracted from the MIRI LRS data in the 5--11 $\mu$m wavelength range can be described with an absorbed PL model with the spectral slope $\alpha=0.963\pm 0.015$, extinction $A_V=3.9\pm 0.2$, and normalization $f_0=59.4\pm 0.5$ $\mu$Jy at $\lambda = 8$ $\mu$m.
    \item The NIRCam F140M and F250M photometric data points, with flux densities $f_\nu =5.1\pm 0.3$ $\mu$Jy and $14.5\pm 0.4$ $\mu$Jy at 1.4 $\mu$m and 2.5 $\mu$m, respectively, agree  very well (within 1$\sigma$) with the extrapolation of this absorbed 
    PL model.
    \item Although the results of Spitzer IRAC photometry at 4.5 $\mu$m and 8 $\mu$m, presented by \citet{2006Natur.440..772W} (and re-analyzed by \citealt{2008ApJ...675..695W}), agree with the JWST spectrum,
    the PL+disk model, suggested by \citet{2006Natur.440..772W}, strongly disagrees with the JWST results.
    \item The extrapolated absorbed PL model also agrees reasonably well with the optical data from \cite{2016MNRAS.458L.114M}. However, there is still strong disagreement with older NIR-optical observations, particularly in the NIR bands, and with  the Spitzer MIPS 24 $\mu$m 
    upper limit.
    \item We favor variability
    as the explanation for these discrepancies, but we cannot exclude the possibility that a contribution from a fallback disk plays a role, at least in the IR part of the spectrum. 
    The origin of the IR-optical emission of 4U\,0142+61 can be tested with nearly simultaneous 
    deep JWST and HST observations across a number of bands in the broad IR-optical range. Monitoring of the source with large ground-based telescopes with simultaneous multi-band imaging capabilities, such as HiPERCAM \citep{2021MNRAS.507..350D}, could also help constrain the variability of the source.
    \item We find that the MIR-NIR spectrum requires a large spectral break ($\Delta \alpha > 1$) to be consistent with the hard power law observed in X-rays, suggesting that the IR and hard X-ray emission are likely being emitted from two different particle populations.
\end{itemize}
Future approved JWST timing observations of 4U 0142+61 will allow us to constrain its pulsed fraction in the IR, help to further elucidate the nature of the IR emission, and help to determine its relationship, if any, to the NIR/optical, and X-ray emission. 

\medskip\noindent{\bf Acknowledgments:}
We thank Sarah Kendrew, Brian Brooks, Bryan Hilbert for their useful advice on JWST data reduction. JH thanks George Younes and Zorawar Wadiasingh for useful discussions related to this project. This work is based on observations made with the NASA/ESA/CSA James Webb Space Telescope. We thank the anonymous referee for carefully reading the manuscript and providing useful feedback. The data were obtained from the Mikulski Archive for Space Telescopes at the Space Telescope Science Institute, which is operated by the Association of Universities for Research in Astronomy, Inc., under NASA contract NAS 5-03127 for {\sl JWST}. These observations are associated with program \#2635 and can be accessed via \dataset[doi:10.17909/1azj-q742]{http://dx.doi.org/10.17909/1azj-q742}. Support for program \#2635 was provided by NASA through a grant from the Space Telescope Science Institute, which is operated by the Association of Universities for Research in Astronomy, Inc., under NASA contract NAS 5-03127. JH acknowledges support from NASA under award number
80GSFC21M0002 and partial support through NuSTAR award NNH21ZDA001N-NUSTAR. This work made use of data supplied by the UK Swift Science Data Centre at the
University of Leicester.

\appendix

\section{Measuring the NIRCam flux densities}
\label{sec:append_A}

\begin{figure*}[h!]
\centering
\includegraphics[width=17cm]{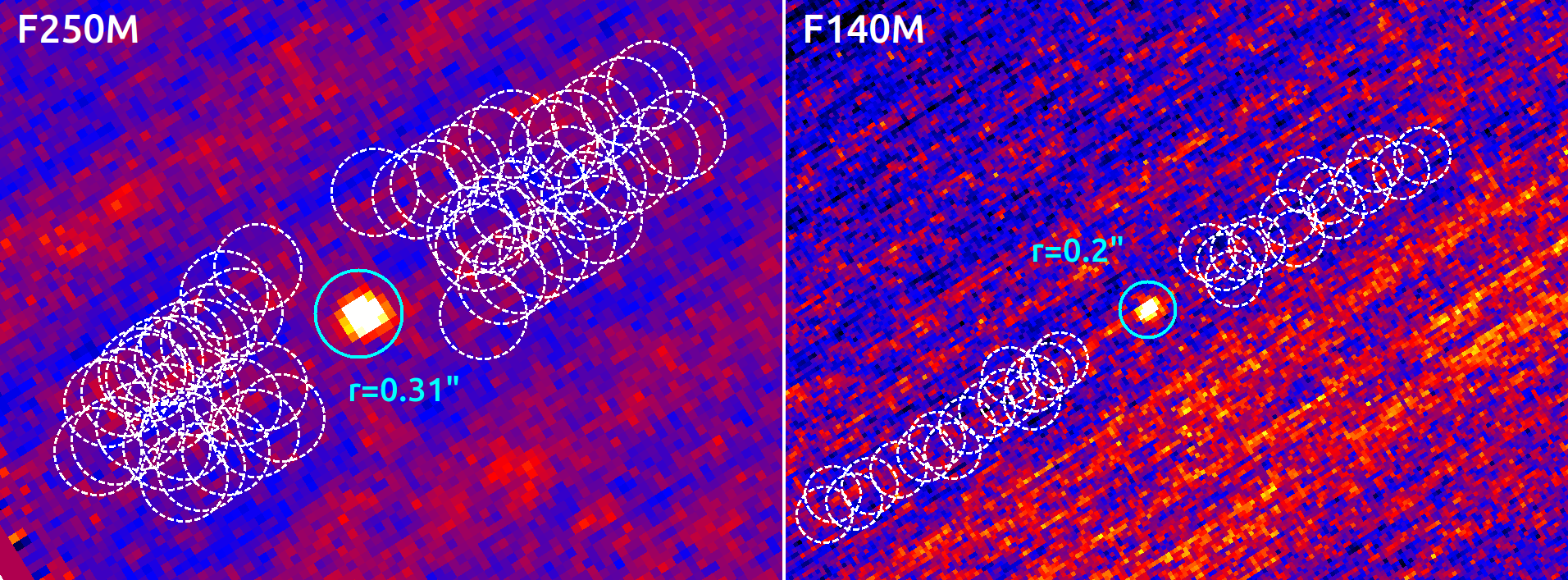}
\caption{The 
NIRCam F250M (left) and F140M (right) images of the magnetar. North is up, East to the left. The 1/f noise pattern is clearly visible in both images. For F250M (F140M), we use 55 (30) white background apertures with the same size as the source aperture in the ``empty aperture'' approach, see text.}
\label{f140f250imas}
\end{figure*}
\begin{figure*}[h!]
\centering
\includegraphics[trim={0 0 0 0},width=8.9cm]{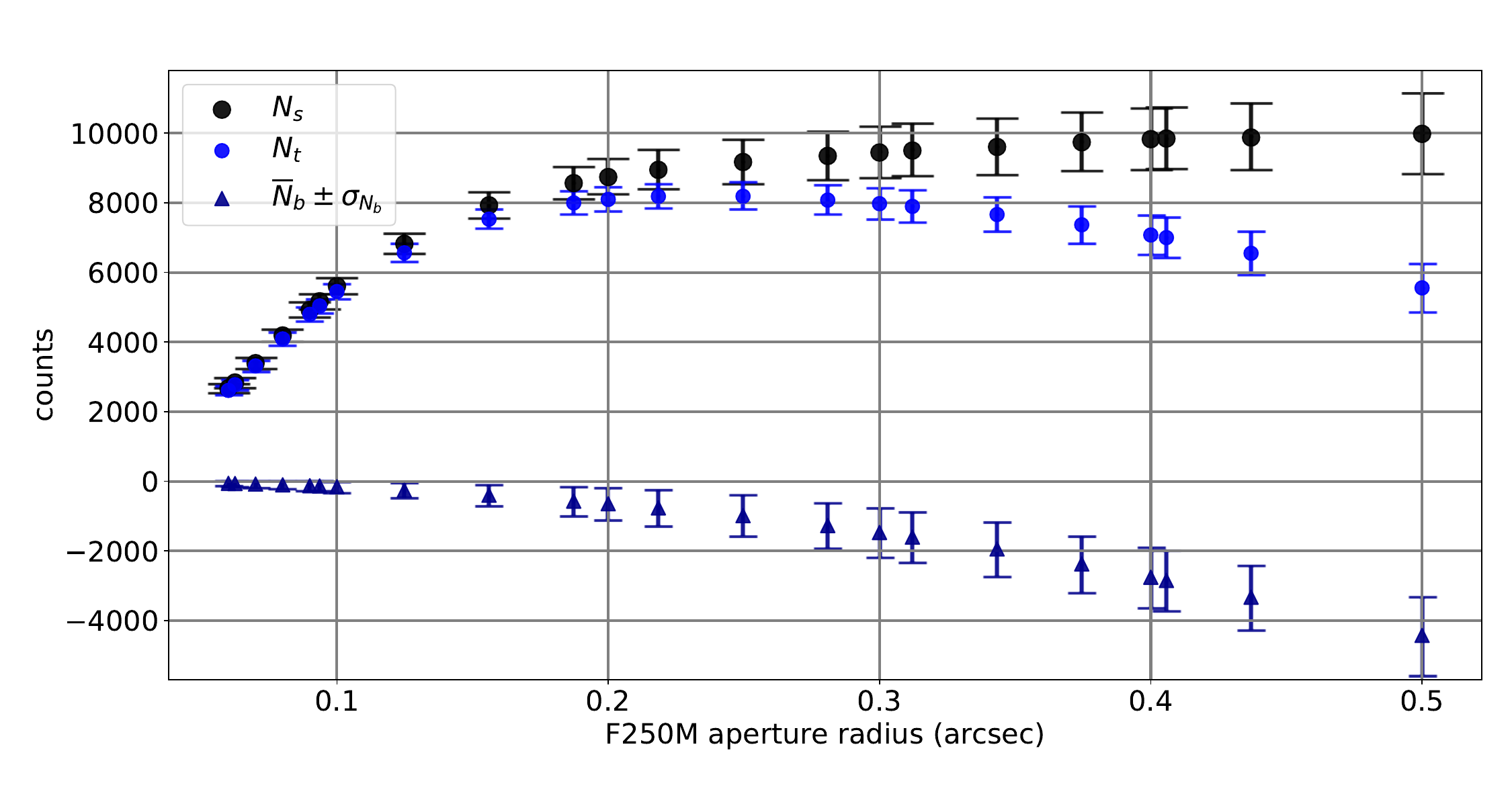}
\includegraphics[trim={0 0 0 0},width=8.9cm]{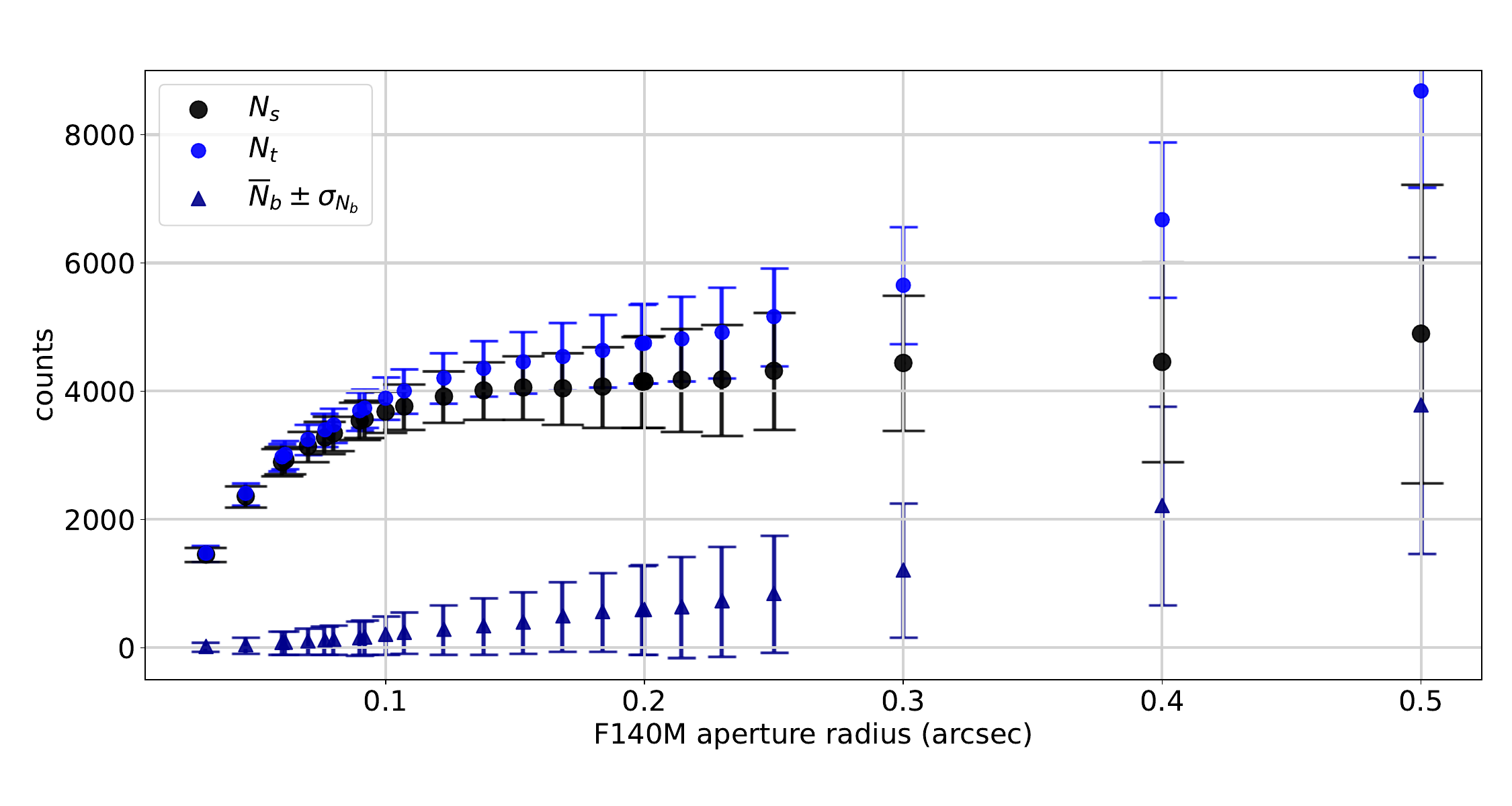}
\caption{Aperture measurements in the NIRCam F250M and F140M filters with total, net, and background counts ($N_t$, $N_s$, $N_b$) as indicated. The background apertures always have the same size as the source apertures. Considering the 1/f noise seen in Figure~\ref{f140f250imas}, and (corrected) flux uncertainties, $0.31\arcsec$ and $0.2\arcsec$ are 
chosen as the optimal apertures for F250M and F140M, respectively.} 
\label{apmeasF250F140}
\end{figure*}

The F250M and F140M images, shown in Figure \ref{f140f250imas}, exhibit strong 1/f noise patterns (see \citealt{2020AJ....160..231S} for additional details).
We use the approach of many ``empty background apertures'' (e.g., \citealt{2014ApJS..214...24S,2022ApJ...924..128A}). By placing our 55 (30) overlapping background apertures in the same 1/f noise bands as the source aperture for F250M (F140M),
we obtained error estimates of the net source counts, $N_s$. The uncertainty of these net counts, $\sigma_{N_s}$, consists of contributions from the net counts themselves (estimated as a Poisson error) and the background noise contribution (the variance $\sigma^2_{N_b}$):
\begin{equation} 
\sigma_{N_s} = {\left( \sigma^2_{N_b} + N_s /g \right)}^{1/2},
\end{equation}
where $g$ is the gain\footnote{ 
See \url{https://jwst-docs.stsci.edu/jwst-near-infrared-camera/nircam-instrumentation/nircam-detector-overview/nircam-detector-performance}},  
which converts the image counts (ADU) into 
detected electrons (see Table \ref{tab:fluxesA}).  

We choose the same size of source and background apertures. The source aperture is centered on the position obtained from a 2D Gaussian fit in the image.
Figure~\ref{apmeasF250F140} shows the measured total counts, median background counts, and net counts as functions of aperture radius.
Aiming to minimize the noise to count ratio and considering the size of the 1/f noise band, we choose $0\farcs31$ (or 5 pixels) and $0\farcs20$ (or $6.5$ pixels) as optimal aperture radii for the F250M and F140M images, respectively. 
Applying the respective (linearly interpolated) encircled energy fractions (versions ETCv2) \footnote{See \url{https://jwst-docs.stsci.edu/jwst-near-infrared-camera/nircam-performance/nircam-point-spread-functions}} 
and the image unit conversion, we calculated the aperture-corrected net source counts and flux densities $f_\nu$ (see Table \ref{nircamphot}). Note that for both filters the uncertainties are mostly
due to the nonuniform 1/f background, i.e., $\sigma_{N_s} \approx \sigma_{N_b}$.
\begin{deluxetable*}{ccccccccccrrr}[h]
\tablecolumns{12}
\tablecaption{Pulsar NIRCam photometry using many background apertures\label{tab:fluxesA}}
\tablewidth{0pt}
\tablehead{
\colhead{Band} & \colhead{$\lambda_{\rm piv}$} & \colhead{$\delta\lambda$} & \colhead{${\cal P}_\nu$} & \colhead{$g$} & \colhead{$t_{\rm exp}$} & 
\colhead{$r_{\rm ap}$} & \colhead{$\phi$} & 
\colhead{${N_{t}}$} & \colhead{${\cal N}_b$} & \colhead{$\overline{N}_b \pm \sigma_{N_b}$} &  \colhead{${N}_{s}$} &
\colhead{${f_{\nu}}$} \\ 
\colhead{} & \colhead{$\mu$m} & \colhead{$\mu$m} & \colhead{$\mu$Jy s/cnt} & \colhead{e$^{-}$/cnt} & \colhead{sec} &
\colhead{arcsec} & \colhead{} & 
\colhead{cnt} & \colhead{} & \colhead{cnt} & \colhead{cnt} &
\colhead{$\mu$Jy} 
}
\startdata
F140M & 1.404 & 0.142 & 0.121 & 2.05 & 115.9 & 0.2  & 0.847 & 4744 & 30 & $ 595 \pm 231$ & $4150 \pm 236$ & $5.1  \pm 0.3$  \\
F250M & 2.503 & 0.181 & 0.152 & 1.82 & 115.9 & 0.31 & 0.857 & 7898 & 55 & $-1602 \pm 241$ & $9501 \pm 251$ & $14.5 \pm 0.4$ \\
\enddata
\tablecomments{
$\lambda_{\rm piv}$ and $\delta\lambda$ are the pivot wavelength and bandwidth of the filter,
${\cal P}_\nu$ is the filter's inverse sensitivity,
$g$ is the gain, 
$t_{\rm exp}$ is the exposure time,
$\phi$ is the encircled count fraction in the source aperture with radius $r_{\rm ap}$,
${N}_t$ indicates the total (source plus background) counts in the source aperture,
${\cal N}_b$ is the number of background apertures considered,
$\overline{N}_b$ and $\sigma_{N_b}$ are the median and variance of these ${\cal N}_b$ background measurements,
${N}_s = (N_t -\overline{N}_b) \pm \sigma_{N_s}$ lists the net (source) number of counts,
${f_{\nu}} = {\cal P}_\nu N_{s} (\phi \, t_{\rm exp})^{-1}$ is the spectral
flux density at the pivot wavelength.}
\label{nircamphot}
\end{deluxetable*}

\end{document}